\documentclass[aps,prl,twocolumn]{revtex4}%,showpacs
\usepackage{epsfig, amsmath, amsfonts, amssymb, bm, graphicx}
\usepackage{color}

\begin{document}

\DeclareGraphicsExtensions{.eps,.EPS,.jpg}
%%%%%%%%%%%%%%%%%%%%%%%%%%%%%%%%%%%%%%%%%%%%%%%%%%%%%%%%%%%%%%%%%%%%%%%%%%%%%%%%%%%%%%%%%%%%%%%%%%%%%%%%%%%%%%%%%%%%%%%%%%%%%%%%%%%%%%%%%%%%%%%%%%%%%%%%%%%%%%%%%%%%%%%%%%%%%%%%%%%%%%%%%%%%%%%%%%%%%%%%%%%%%%%%%%%%%%%%%%%%%%%%%%%%%%%%%%%%%%%%%

\title{Measuring densities of cold atomic clouds smaller than the resolution limit.}
\author{A. Litvinov$^1$, P. Bataille$^1$, E. Mar\'echal$^{1,2}$, P. Pedri$^1$, O. Gorceix$^1$, M. Robert-de-Saint-Vincent$^1$, B. Laburthe-Tolra$^1$.}
\affiliation{$^1$Laboratoire de Physique des Lasers, Universit\'e Sorbonne Paris Nord F-93430
Villetaneuse, France and LPL CNRS, UMR 7538, F-93430 Villetaneuse, France \\$^2$Laboratoire de Physique et d'Etude des Mat\'eriaux (LPEM), CNRS, Universit\'e Paris Sciences et Lettres}

\begin{abstract}

We propose and demonstrate an experimental method to measure by absorption imaging the size and local column density of a cloud of atoms, even when its smallest dimension is smaller than the resolution of the imaging system. To do this, we take advantage of the fact that, for a given total number of atoms, a smaller and denser cloud scatters less photons when the gas is optically thick. The method relies on making an ansatz on the cloud shape along the unresolved dimension(s), and on providing an additional information such as the total number of atoms.
We demonstrate the method on \textit{in-situ} absorption images of elongated $^{87}$Sr Fermi gases. We find significant non-linear corrections to the estimated size and local density of the cloud compared to a standard analysis. This allows us to recover an un-distorted longitudinal density profile, and to measure transverse sizes as small as one fourth of our imaging resolution. The ultimate limit of our method is the wavelength that is used for imaging.

\end{abstract}

\date{\today}
\maketitle
\vspace{0.5cm}

%\section{Introduction}
The cold atoms community typically uses absorption imaging to explore the physics of dense clouds of atoms. In many cases, especially in the case of \textit{in situ} observations to investigate $e.g.$ the equation of state of quantum fluids \cite{Shin2008, ho2009, Nascimbene2010, Yefsah2011} or 1D gases \cite{cazalilla2011}, these objects are extremely small and sometimes even below the resolution limit of the imaging system. More generally, quantum gases reveal interesting phenomena in small, local features: vortices \cite{vortex} whose size is set by condensates' healing length and typically cannot be resolved \textit{in situ}, density fluctuations \cite{fluctuations}, Wannier functions in optical lattices, etc. Resolving such structures is a difficult but rewarding problem that prompted important technical developments, using for example high-resolution objectives in the quantum gas microscope approach \cite{gross2017} (which are nevertheless still limited to the diffraction limit), the newly demonstrated quantum gas magnifier \cite{Asteria2021}, super-resolution imaging \cite{Subhankar2019, Donald2019}, scanning probes using electrons \cite{wurtz2009} or ions \cite{Veit2021}. 

Here we focus on the \textit{in-situ} imaging of small and dense objects using standard absorption imaging. In practice, trapped gases can be both extremely absorbing and smaller than the imaging resolution, which constitutes a severe difficulty for image analysis due to $e.g.$ total light absorption \cite{Reinaudi2007} or diffraction of the optical fields used for imaging. We nevertheless show that it is still possible to accurately measure the size and local density of an object, even when this object is smaller than the imaging resolution. 

The decrease in light intensity induced by propagation through an atomic sample can be linked to the column density thanks to the Beer-Lambert law.
However, for objects smaller than the imaging resolution and when light absorption is strong, the information on the column density is partially lost. Indeed, the Beer-Lambert law is non linear and cannot be averaged over the imaging resolution.
Crucially, the average number of absorbed photons per atom depends on the cloud size since atoms in a smaller and therefore optically thicker medium are exposed to a reduced average light intensity. Our main idea is to take advantage of this to reconstruct information on size and local density at scales below the imaging resolution.

\begin{figure*}[t!]
    \centering
    \includegraphics[width=1.\textwidth]{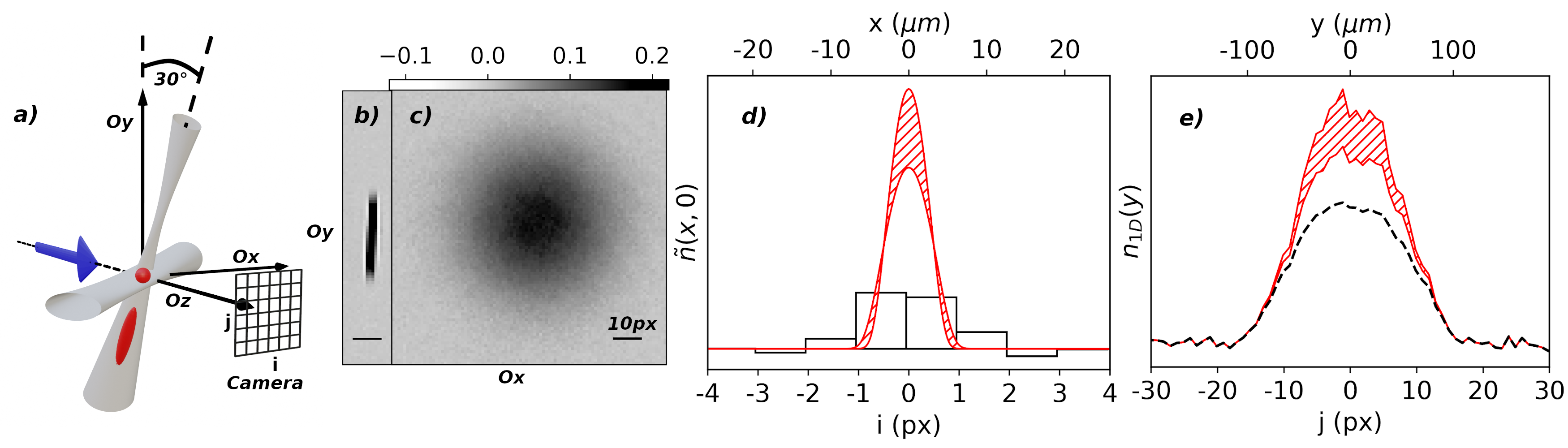}
    \caption{Recovering density profiles from distorted absorption images of tightly confined gases. a) Geometry. b) Absorption image of a gas tightly confined and expanding in one dimension. c) Time of flight absorption image when the gas is released in all three directions. d) Cross section along the short axis of the column density shown in b), at peak value. The raw data (column density deduced from the pixelated optical depth) shows diffraction fringes locally leading to non-physical negative optical depth (black squares). Our method does not use the optical depth, but rather estimates density by counting missing photons and assuming a Gaussian shape to the cloud in its shortest dimension. The result of our method is shown by the red solid lines, delimiting a confidence interval (red hashes) resulting from experimental fluctuations. The raw data underestimates the atom number by a factor 1.5, and the peak column density by a factor of roughly 4 in this case, while our method allows to infer an object size of only 0.25\,px (pixel size). e) Linear density along the long axis \textit{(Oy)}, $n_{1D}(y)$ deduced either by integrating the pixelated optical depth along the short axis \textit{(Ox)} (black dashes), or by our method (red solid lines, delimiting the confidence interval).}
    \label{fig1}
\end{figure*}

In the first section of this paper, we explain our image processing, which estimates the size of atomic clouds by using the non linearity of the Beer Lambert law and an additional information. This latter,  we here take to be an independent measurement of the total number of atoms. In the second section, we use this method to analyze images of ultra-cold strontium clouds at temperature ranging from $0.14\,T_F$ to $1.6\,T_F$ (where $T_F$ is the Fermi temperature). We focus on cases where the ultra-cold gases have expanded while being strongly confined in two dimensions, which results in a very elongated cloud with a very small transverse dimension due to confinement, and a longitudinal profile that reproduces the initial momentum distribution. This very elongated cloud, and in particular the measurement of its transverse size, is our subject of study. In practice, using the averaged Beer Lambert law results in a loss of information on this transverse size (which is smaller than the resolution limit of our imaging system), and a strongly distorted shape along the longitudinal axis. We show that our method enables us to deduce the transverse size, which in practice was down to 0.25 times the imaging resolution at the lowest temperature. This transverse size is in good agreement with theoretical predictions; the longitudinal density profile deduced from our method is also in agreement with the expected Fermi distribution.

\section{Beer-Lambert Law for objects below imaging resolution}

We first describe our approach to analyze absorption images of atomic clouds smaller than the imaging system resolution. We will focus on a situation where the profile of the column density is smaller than the resolution limit in one direction only. The approach can be generalized to structures that are smaller than the imaging resolution in the two dimensions perpendicular to the propagation axis of the imaging laser beam.

Under standard conditions of imaging, the propagation along $Oz$ of resonant light at intensities $I(x,y,z)$ below the saturation intensity, through an atomic cloud of density $n(x,y,z)$, follows the Beer-Lambert law

\begin{equation}
    \frac{dI(x,y,z)}{I(x,y,z)}=-n(x,y,z)\sigma_0 dz
\label{beer-lambert}
\end{equation}

where $\sigma_0=3\lambda^2/2\pi$ is the resonant light absorption cross-section for a single atom. $\lambda$ is the photon wavelength. We will consider the situation where the longitudinal size of the object $l$ along the direction of the imaging laser beam is smaller than the Rayleigh length associated with its transverse size $\sigma$, $i.e.$  $l< \pi \sigma^2 / \lambda$. In this regime, diffraction of the imaging beam as it propagates through the cloud can be neglected over the distance $l$, such that the intensity variations along $x,y,z$ are independent. Then Eq.(\ref{beer-lambert}) can be integrated to provide a measurement of the optical depth (OD):
\begin{equation}
    OD(x,y)=-\log \frac{I(x,y)}{I_0(x,y)}=\sigma_0\int_{-\infty}^{\infty} n(x,y,z)dz
\label{opticaldepth}
\end{equation}
where $I_0(x,y)$ is the transmitted light intensity without atoms and $I(x,y)$ the transmitted light intensity with atoms. $OD(x,y)/ \sigma_0$ thus provides the column density $\int_{\mathbb{R}}n(x,y,z)dz \equiv  \Tilde{n}(x,y)$.

Experimentally, information on light intensity is limited by the imaging resolution, set by $e.g.$ the detector pixel size, aberrations,  or the diffraction limit. We thus consider the collected power over a size $a$, at position $\left\{ x=ia,y=ja  \right\} $, characterized by the indices $(i,j)$ 

\begin{equation}
    %\bar{I}(i,j)=\frac{1}{a^2}\iint_{ia,ja}^{(i+1)a,(j+1)+a} I(x,y)dxdy
    P(i,j)=\iint_{D_{i,j}} I(x,y)dxdy
\label{pxtopx}
\end{equation}
where $D_{i,j}=  \left[ \left\{ ia,(i+1)a  \right\} \times \left\{ ja,(j+1)a  \right\} \right]$ is the domain of integration. It is possible to extract density information from 
$\frac{P(i,j)}{P_0(i,j)}$ (where $P_0(i,j)$ is the collected power without atoms) and Eq.\,(\ref{opticaldepth}) when $I(x,y)$ varies slowly over the distance $a$, or when absorption is negligible. However, in general, and in the present situation,

\begin{equation}
    \log \left( \frac{P(i,j)}{P_0(i,j)} \right) \neq  \frac{1}{a^2} \iint_{D_{i,j}} \log \left( \frac{I(x,y)}{I_0(x,y)} \right)  dxdy
\label{condition}
\end{equation}
such that only measuring $P(i,j)$ and $P_0(i,j)$ is insufficient to provide information on either the peak or the averaged column density. 

Nevertheless, for a given total atom number, the total absorbed light power strongly depends on the extension of the imaged object in the imaged plane when absorption is strong, irrespective of the imaging resolution. This is due to a shadowing effect in which the first atoms met by the imaging light reduce the light intensity for the subsequent atoms. This is strongest for small size samples because of increased density. While this effect is well captured by the Beer-Lambert law, Eq. (\ref{condition}) indicates deviations when information is derived from images that are pixelated or blurred at the scale of the sample size.  Then, the number of absorbed photons still depends on the size of the cloud, and, as we shall now see, it is sill possible to deduce this size and the local density if one provides independent information, such as the total atom number.

In our experiment, the column density  varies rapidly along the short axis $Ox$ and slowly along $Oy$, see Fig. \ref{fig1}. We will now relate the total number of absorbed photons integrated along a pixel line parallel to $Ox$ to the local column density. We define:

\begin{equation}
    R_{ph}(j)=\sum_{i} \frac{P_0(i,j)-P(i,j)}{P_0(i,j)}
\label{missingphotons}
\end{equation}

For brevity, we introduce $\delta I(x,y) = I_0(x,y)-I(x,y)$. We then relate the pixelated information 
to local intensities $I(x,y)$ using Eq.\,(\ref{pxtopx}), and then use Eq.\,(\ref{opticaldepth}). 
\begin{eqnarray}
R_{ph}(j) &=& \sum_i \frac{1}{P_0(i,j)} \iint_{D_{i,j}} dx dy  \,\delta I(x,y) \nonumber \\
~&=& \sum_i \iint_{D_{i,j}} dx dy  \frac{I_0(x,y)}{P_0(i,j)} \nonumber\\
~&~& \times \Big(1-\exp(-\sigma_0\int_{\mathbb{R}}n(x,y,z)dz)\Big)
\label{toto}
\end{eqnarray}

The discrete indices $(i,j)$ and the size $a$ can be matched to physical pixels of the camera, or to effective pixels after binning. In any case, 
Eq.\,(\ref{toto}) is valid provided the incident intensity in the atomic plane, imaged to form $I_0(x,y)$ on the camera, is homogeneous at the spatial scale of the resolution limit of the imaging system. 

To simplify Eq.\,(\ref{toto}), we note that $\frac{I_0(x,y)}{P_0(i,j)} = \frac{1}{a^2}$, and that $\sum_i \int_{ia}^{(i+1)a}dx [..] = \int_{\mathbb{R}} dx [..] $ as the integrand $[..]$ is zero far away from the cloud. Furthermore, in our case, the atomic density along $Oy$ varies slowly over a pixel length, such that we can also replace $\int_{ja}^{(j+1)a}dy f(y)$ (where $f$ is the integrand over $y$ in Eq.\,(\ref{toto})) by $a f(ja)$. 

Our idea is to introduce an ansatz on the local column density profile along $Ox$, which for simplicity is taken 
as Gaussian:
\begin{equation}
    %n(x,y,z) = e^{-\frac{x^2}{2\sigma_x^2}} n(x=0,y,z),
    \Tilde{n}(x,y) = e^{-\frac{x^2}{2\sigma_x^2}} \Tilde{n}(0,y),
\label{ansatzGauss}
\end{equation}
Then, in the integration along $x$ of Eq.(\ref{toto}) we apply a change of variable $u=x/\sigma_x$. We obtain:

\begin{eqnarray}
    R_{ph}(j) &=& \frac{\sigma_x}{a} \int_\mathbb{R}\Big(1-\exp \Big(-\sigma_0 \Tilde{n}(0,aj)e^{-\frac{u^2}{2})}\Big)du \nonumber \\
    ~&\equiv& \frac{\sigma_x}{a} F\big(\sigma_0 \Tilde{n}(0, aj)\big),
    \label{transferF}
\end{eqnarray}
where we defined the transfer function $F$. 

% We point out that, in case the number of atoms is not easily available, another piece of information can be used to retrieve this transverse size, such as for example the longitudinal cloud shape along $Oy$, that should for example obey Boltzmann statistics and be Gaussian at high temperature.

The total number of atoms $N_{at}$ can be computed according to the definition of the column density and to Eq.(\ref{ansatzGauss}), and related to $R_{ph}$ using Eq.(\ref{transferF}):

\begin{equation}
    N_{at}=\frac{\sqrt{2\pi} a \sigma_x}{\sigma_0}\sum_j F^{-1}\big(\frac{a}{\sigma_x}R_{ph}(j)\big)
\label{Nat}
\end{equation}

To obtain these equations, four assumptions have been made. First, the density along the short axis $Ox$ is assumed to be a Gaussian of width $\sigma_x$. Second, the column density $\Tilde{n}(x=0, y)$ along the long axis $Oy$ is quasi-uniform along the width of a pixel. Third, the illumination is homogeneous at the scale of the resolution limit. Finally we also assumed that diffraction of the probe light by the atomic sample can be neglected over the depth of the sample along $Oz$, so that the Beer-Lambert law can be applied locally. 

Both Eq.(\ref{transferF}) and Eq.(\ref{Nat}) give the possibility to reconstruct the actual density profiles even when the Beer-Lambert law cannot be directly used on the raw data. Indeed, $F$ is non linear so that a given density profile $\Tilde{n}(0, aj)$ results in distorted profiles $R_{ph} (j)$ that parametrically depend on $\sigma_x$. Therefore, we can in principle retrieve the transverse size $\sigma_x$ from Eq.(\ref{transferF}) when the longitudinal density profile is known a priori. This is the case for example in our experiment  at high temperature, where  the longitudinal cloud shape along $Oy$ should  obey Boltzmann statistics and be Gaussian. Alternatively Eq.(\ref{Nat}) simply relates the free unknown parameter $\sigma_x$ to the total number of atoms, which is useful provided this number can be measured independently. 

In what follows, we obtain $N_{at}$ by an independent measurement on expanded clouds for which the gas is larger than the resolution in 3D, so that the usual analysis holds and $N^{3D}_{at} = \frac{a^2}{\sigma_0} \sum_{i,j} OD^{3D}(i,j)$ ($OD^{3D}$ is the optical depth for the images with 3D expansion). We  point out that the estimate of $\sigma_x$ that we thus perform is independent from uncertainties in $\sigma_0$ (uncertainties that can arise for example for atoms with an hyperfine structure due to optical pumping effects during the imaging pulse). Indeed, setting $N_{at}=N^{3D}_{at}$, from the explicit expressions of $N_{at}$ and $N^{3D}_{at}$ if follows immediately that $\sigma_x$ is independent of $\sigma_0$.

\section{EXPERIMENTAL RESULTS}

In our experiment, whose setup is detailed in \cite{bataille2019}, the $^{87}$Sr cold atomic sample is obtained after evaporation to reach regimes from T$\simeq$\,2\,$T_F$ down to T$ \simeq 0.15\,T_F$. The optical dipole trap is made of an horizontal anisotropic laser beam crossed by a second laser beam at 30 degrees from vertical, see Fig.1.a. To produce a tightly confined gas in one dimension, we switch the horizontal beam off. This allows for an expansion of the gas channeled by the second laser beam, for times ranging from 0.1\,ms to 20\,ms. We then take an absorption image, see Fig.1.b, using a pulse of circularly polarized resonant light whose intensity is about 50 times below the saturation intensity, and a non-magnifying imaging system that uses a telescope configuration to approximately conjugate the atomic plane to a CCD chip with 6.5\,$\mu$m-wide pixels. The telescope is made of two achromatic lenses of local length $f=150$\,mm, and diameter 50 mm. The opening diameter is set to 30 mm to reduce geometrical aberrations, and the diffraction-limited resolution corresponds to 2.8\,$\mu$m. However, in practice, the resolution of our imaging system is limited by the pixel size of the CCD chip.  Furthermore, as a function of time, the cloud is positioned at different depths $z$ along the imaging axis, due to the orientation of the guiding beam (see Fig 1.a). This results in images that can be slightly out-of-focus. To independently calibrate the number of atoms, we release the gas by switching off both trapping lasers and take absorption images after a three dimensional free flight, see Fig.1.c. 

As can be seen in Fig.1.b, images of tightly confined gases are affected by pixelation and diffraction fringes perpendicular to the elongated direction. Fig.1.d shows a cut through those fringes, which are visible because the object is slightly out of focus. The only diffraction signature is along the transverse axis, and arises because of fast variations of the atomic density along this axis. 
%As light is redistributed between pixels only along this direction, with slow variations along $Oy$ of both light and atomic density, 
%Due to slow variations along $Oy$ of both light and atomic density, 
Thus the light is redistributed along this direction only. 
Therefore, we can measure $R_{ph}(j)$ on camera pixel lines without binning, even when the pixel size $a$ is smaller than the actual optical resolution on out-of-focus clouds. Using Eq.\,(\ref{transferF}) we deduce the transverse size and the local density profile of the gas, also shown in Fig.1.d.
%Thus the light is redistributed along this direction only. 
%Therefore, we can sum the missing photons along each pixel line transverse to the elongated direction to deduce the local number of absorbed photons, and use Eq.(\ref{transferF}) to deduce the transverse size and the local density profile of the gas, also shown in Fig.1.d. 

Our approach thus recovers the local peak density from which we also infer the actual density profile along the elongated axis. As shown in Fig.1.e the linear density derived by integrating the pixelated optical depth has a distorted shape compared to the linear density, as recovered by our method. As far as the low density part is concerned the reconstructed profile matches that deduced from integrating the pixelated optical depth, as expected for low absorption. In contrast, the high density region, corresponding to larger absorption, is underestimated when using the pixelated optical depth. 

\subsection{Analysis of the sub-resolution transverse size}

\begin{figure}[t!]
    \centering
    \includegraphics[width=0.5\textwidth]{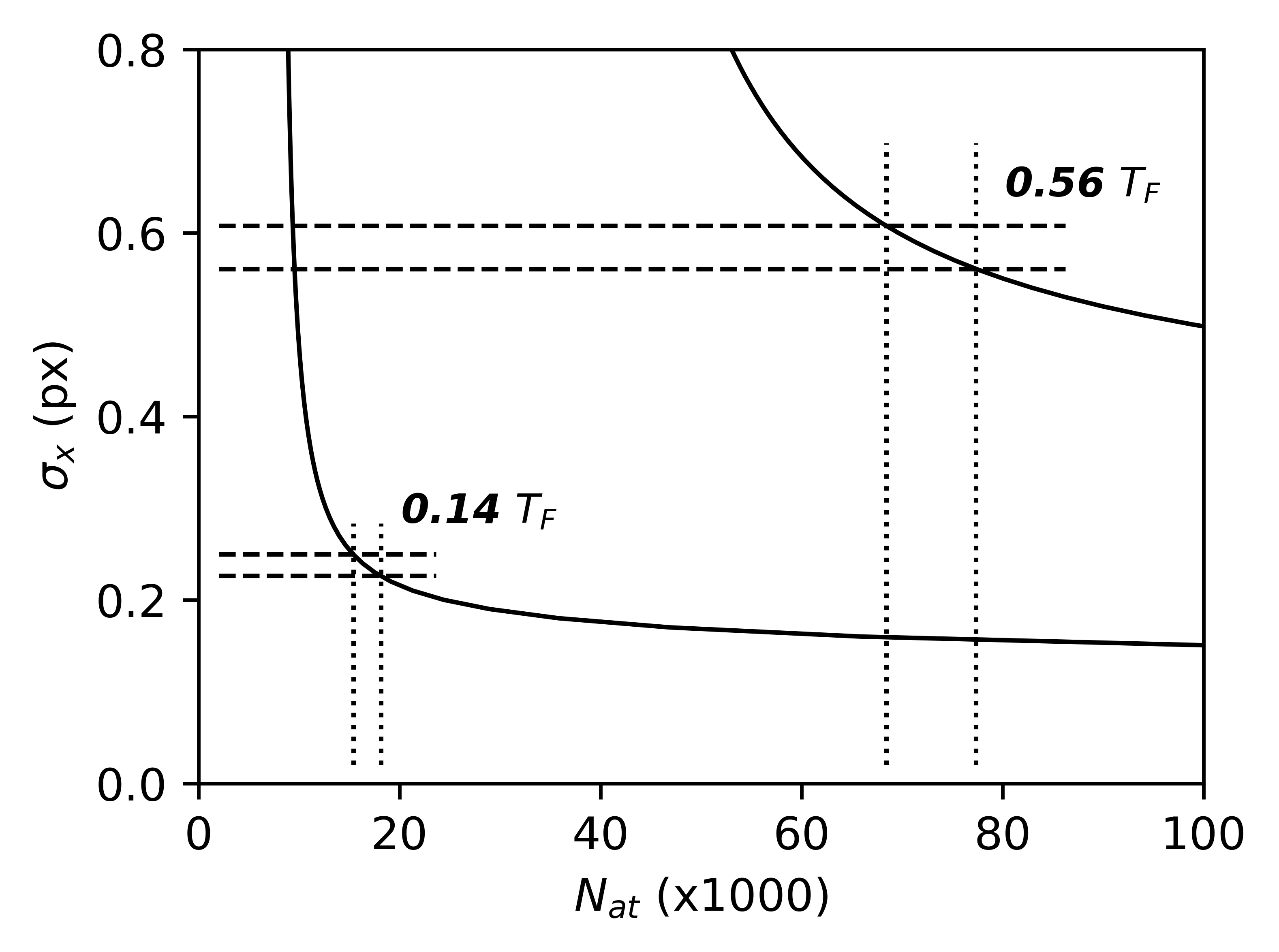}
    \caption{Determination of $\sigma_x$ based on the independently measured total atom number $N_{at}$, for two images (at 0.14 $T_F$ and 0.56 $T_F$) whose peak optical depths significantly differ. The black line is the inverse of Eq.(\ref{Nat}) for a specific $R_{ph}$, \textit{i.e.} a single image. The atom number confidence intervals (delimited by the dotted lines) are due to atom number fluctuations from shot to shot, and are the dominant cause for the $\sigma_x$ confidence intervals (delimited by the dashed lines). Qualitatively, a low optical depth results in a large slope, and thus large $\sigma_x$ uncertainty.}
    \label{fig2}
\end{figure}

\begin{figure}
    \centering
    \includegraphics[width=0.5\textwidth]{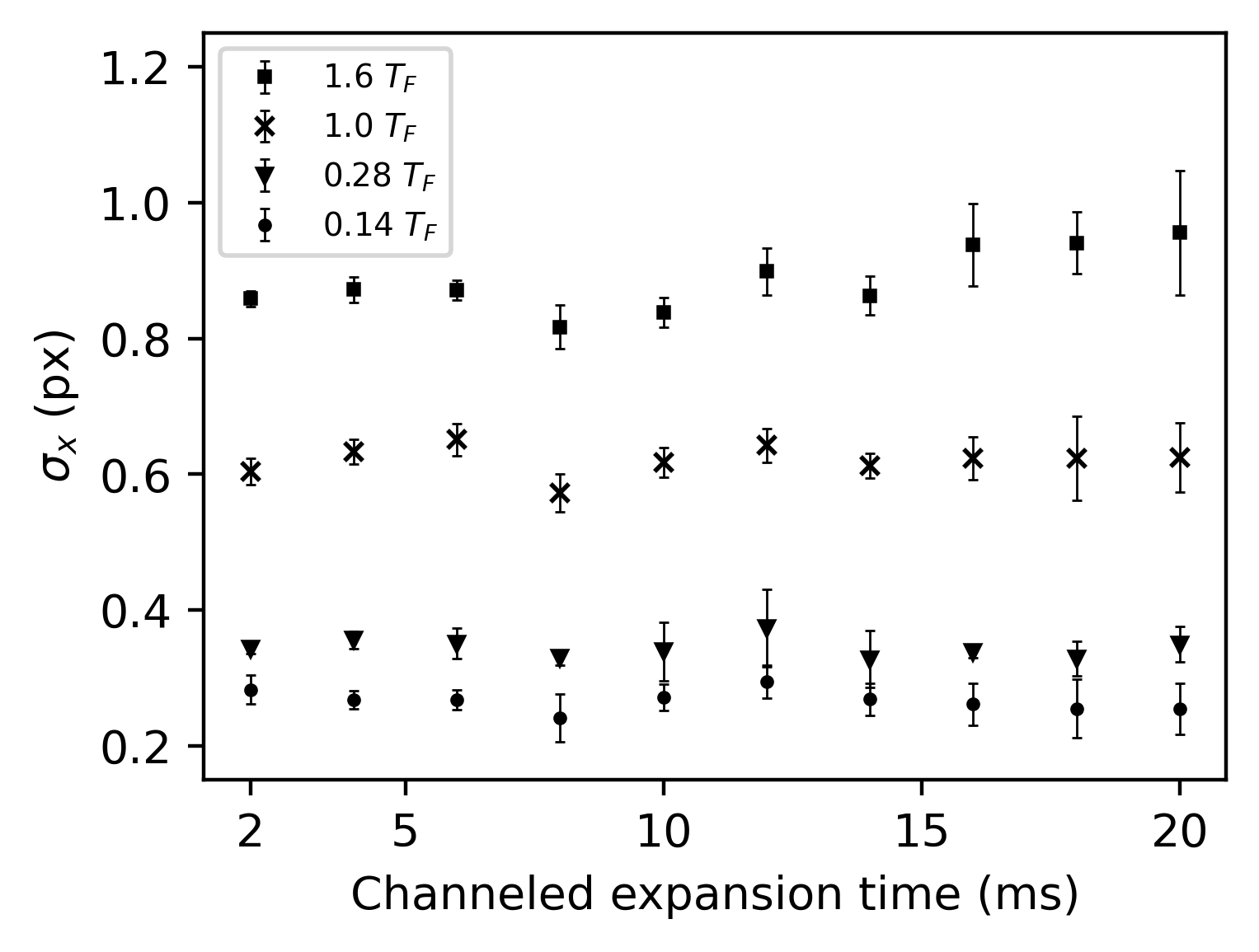}
    \caption{Time dependence of $\sigma_x$ from 2\,ms to 20\,ms channeled expansion, at different degeneracies. Each point aggregates typically 10 images, with error bars that show the standard deviations between the individual measurements. Those standard deviations match the individual picture confidence interval (see Fig.\,\ref{fig2}), and thus can be attributed to mostly measurement uncertainty. The degeneracy is deduced from the reference images.}
    \label{fig3}
\end{figure}

We now focus on our measurements of $\sigma_x$. Fig.\,\ref{fig2} illustrates how the atom number from reference images is used to recover $\sigma_x$. For a given picture of the elongated cloud, we compute $R_{ph}(j)$, we plot the number of atoms $N_{at}$ deduced from Eq.(\ref{Nat}) as a function of an assumed value of $\sigma_x$. From the reference measurement of $N_{at}^{3D}$ and its uncertainty associated with shot to shot fluctuations, we deduce the corresponding value and confidence interval of $\sigma_x$ that we will present on the following figures.

On Fig.\,\ref{fig3} we observe that $\sigma_x$ does not vary much with the guided time of flight (its variations are similar to the approximately $10\%$ standard deviations of the data). Such a small variation is expected because the gas evolves while being confined in the guiding beam with a transverse confinement frequency which we estimate to vary from 210\,Hz at the initial position to 145\,Hz after 20\,ms of guided fall. Indeed, if we assume that the transverse degrees of freedom follow adiabatically such an evolution, the decrease in trapping frequency corresponds to an increase of only $20\%$ in the transverse size. Our ability to measure $\sigma_x$ for a large range of expansion times illustrates the robustness of the method as, from short to long time of flight, the peak optical density varies by a factor 14 because of the longitudinal expansion.
Furthermore, the method provides accurate estimates of $\sigma_x$ even for short expansion times (2\,ms), which indicates that the longitudinal profile varies slowly enough for the approximation made to derive Eq.(\ref{transferF}) to remain valid. 

\begin{figure}[t!]
    \centering
    \includegraphics[width=0.5\textwidth]{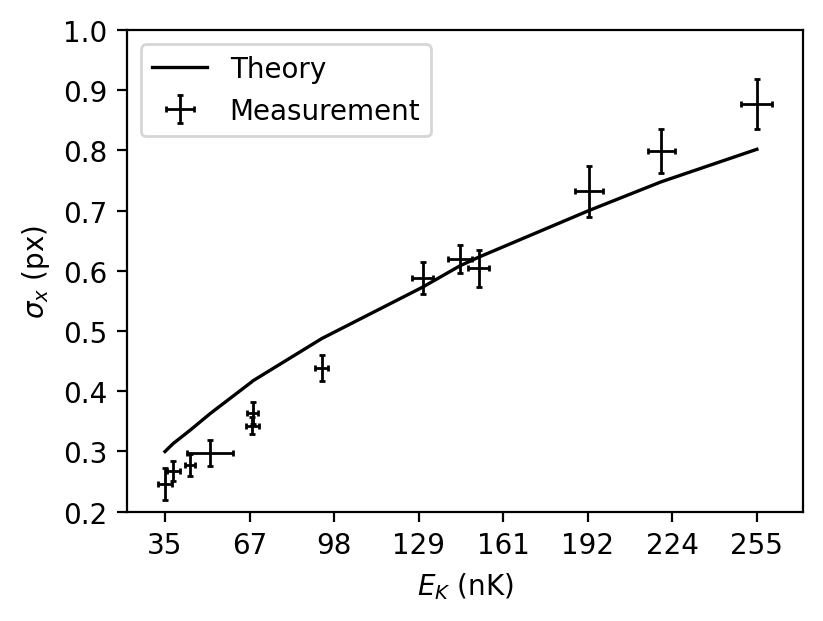}
    \caption{Dependence of $\sigma_x$ as a function of kinetic energy along one direction. The kinetic energy is deduced from the reference images. The black line shows the expected size according to equipartition of energy. The error bars show the standard deviations, aggregating data for all channeled expansion times.}
    \label{fig4}
\end{figure}

As shown in Fig.\,\ref{fig4}, $\sigma_x$ depends on the temperature and reaches 0.25 times the pixel size at lowest temperatures, $T\simeq30\,$nK at $0.14\,T_F$.  This value, $i.e.$ $1.6 \, \mu$m, is smaller than the typical distortions introduced by diffraction of the imaging beam and out-of-focus measurements. It is also smaller than the ultimate resolution of our imaging system (set by the pixel size).  

To compare our measurements to theoretical expectations, we use the reference images of 3D expansions, and we measure the mean kinetic energy along the longitudinal axis $E_K$. Based on the equi-partition of energy, and neglecting inter-atomic interactions, we expect $\frac{1}{2}m\omega^2 \sigma_{x_{th}}^2 = E_K$.  In Fig.\,\ref{fig4}, we thus compare the measured dependence of $\sigma_x$ as a function of the energy per atom to the expected values of $\sigma_{x_{th}}$, using the independently measured initial trap confinement frequency. We find an overall good agreement with this rough theoretical model without free parameter, with discrepancies smaller than 15 $\%$ throughout the curve. We point out that, based on the resolution of the Boltzmann equations \cite{Menotti2002}, we expect interaction effects on the measurement of $E_K$ during the expansion to be of the order of 8 percent at the lowest temperature given our experimental parameters \cite{Sonderhouse2020}, while the interaction effects on $\sigma_x$ are of the order of 1 percent. 

\subsection{Longitudinal profile}

\begin{figure}[t!]
    \centering
    \includegraphics[width=0.5\textwidth]{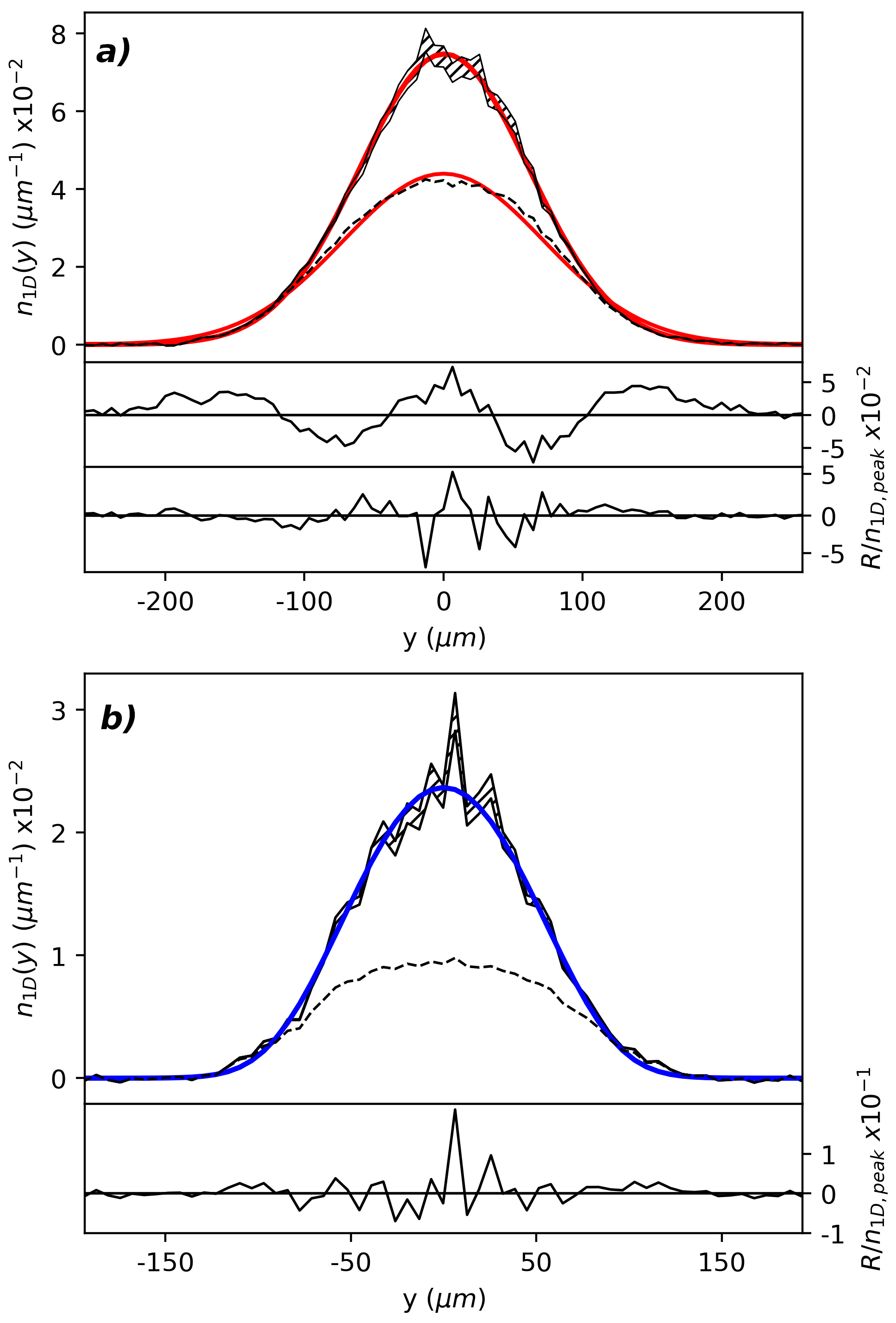}
    \caption{Density profiles along the long axis \textit{Oy}. The black hashed surface is the confidence interval on the measured $n_{1D}(y)$. The dashed line corresponds to the density deduced from the pixelated optical depth. a) Non degenerate gas at 1.6\,$T_F$ after a 10 ms guided expansion. The red lines are Gaussian fits for the linear density deduced from the pixelated optical depth (upper normalized residual) and for $n_{1D}(y)$ (lower normalized residual). The residuals show that the raw data differs significantly from the expected Gaussian shape of a thermal gas. On the contrary, our method recovers this expected Gaussian shape. b) Degenerate gas at 0.21 $T_F$ after a 18 ms guided expansion. In blue, the expected density profile (with no free parameter) matches the profile recovered by our method, as demonstrated by the residuals shown below.}
    \label{fig5}
\end{figure}

In order to further verify that our results match the ones that are expected from the theoretical point of view, we now turn to the longitudinal profiles that we obtain, and compare them to theory. In Fig.\,\ref{fig5}, we present both the longitudinal density profile deduced from the pixelated optical depth, and the 1D density profile $n_{1D}$ deduced from our method. 

Fig.\,\ref{fig5}.a shows a non degenerate gas at $1.6\,T_F$ after a channeled expansion of 10 ms. For such a long expansion, the longitudinal profile should reflect the momentum distribution along this axis before expansion. Both profiles are fitted by Gaussian distributions since at this temperature the momentum distribution is almost perfectly described by Boltzmann statistics. The residuals to the fits show that the 1D profile deduced from the pixelated optical depth (upper residuals) systematically distorts the shape of the cloud, while our reconstruction method (lower residuals) naturally recovers the expected Gaussian profile. 

As shown in Fig.\,\ref{fig1}.a, this expansion is performed along an axis that makes an angle with the imaging plane, which defines a parallax for our measurements. In addition, during the expansion, the atoms feel a position-dependent potential due to the divergence of the Gaussian laser beam. In practice this results in an anti-confinement. The parallax and anti-confinement can both be measured by comparing the size of the expanding cloud to its temperature, and measuring the trajectory of the center of mass. For the analysis shown in Fig.\,\ref{fig5}, the parallax and anti-confinement simply result in a single multiplicative correction parameter C to the cloud size for a given expansion time.

Fig.\,\ref{fig5}.b shows the linear density of a degenerate gas at $0.21\,T_F$. We compare it to the expected density profile of a degenerate Fermi gas expanding in 1D. There are no free parameters since this density profile only depends on the number of atoms, the temperature and degeneracy, which are all measured from the reference images, and $C$, measured from Fig.\,\ref{fig5}.a. Obviously, the linear density profile from pixelated data is in severe disagreement with the theoretical model, unlike the reconstructed profile. The agreement with this latter profile is good throughout the curve as can be seen from the residuals. However, for the lowest momenta, we also observe a strong noise that affects the reconstruction. These irregularities are due to the strong non-linearity of the transfer function $F$ at high density. This non-linearity may also bias an initially symmetric noise. This constitutes a limitation to our reconstruction scheme when the source of the noise is purely from technical origin; however, it also signals that our method could unveil density fluctuations that would otherwise be washed out by the limited optical resolution. 

\subsection{Discussion and conclusion}

We introduced a novel method to analyze absorption images and to measure density profiles at scales below the imaging resolution limit that may occur due to $e.g.$ pixelation, distortion of the images caused by aberrations, and the diffraction by the numerical aperture. These limitations are well known in the community and make the analysis of very small objects particularly difficult. Here we go beyond the approach performed in \cite{bouchoule2006, armijo2010} that corrects the absorption by a simple numerical factor. Since the deviation to the Beer-Lambert law is non-linear, we find that it is important to follow such a procedure as presented here, in order to recover the exact local density and an un-distorted density profile, even along the direction in which the cloud is larger than the imaging resolution. 

In our experiment, we characterize objects as small as 0.25 times our pixel size, despite diffraction fringes due to out of focus imaging  as large as two pixels - these objects are thus significantly below our resolution limit. Comparisons with theoretical models of the cloud shape with no free parameter confirm the validity of these results. We have further checked our method by testing its sensitivity to the ansatz that is chosen for the transverse profile. We found that using a Fermi distribution in the transverse profile barely modifies our results. This good agreement relies on a certain similarity between the actual density profile and the chosen ansatz. However, our method would not be able to reveal \textit{unexpected} features in the shape along the non-resolved direction, since different density shapes could lead to the same number of scattered photons.  

We point out that although our method allows retrieving features below the imaging resolution, its fundamental limit is $\simeq \lambda$, as $\lambda$ sets the scattering cross section of light by the atoms. As a consequence, an object whose size would be below $\lambda$ would affect light propagation identically irrespective of its actual size. In addition to this limit on the transverse size $\sigma$, there is also a limitation associated with the depth $l$ of the object along the axis of the imaging beam. For our equations to hold, it is necessary that the diffraction of the light field can be neglected over the distance $l$. Therefore is is important that $l< \pi \sigma^2 / \lambda$.

Finally, we point out that our method could easily be generalized to study features smaller than the imaging resolution within clouds otherwise larger than the imaging resolution. For example, using the appropriate ansatz, we have in mind to measure the size of vortex cores within a 2D or a 3D superfluid, that are typically below the imaging resolution, or other hydrodynamic structures such as solitons \cite{Denschlag2000}. 

\acknowledgements Acknowledgements: This research was funded by the Agence Nationale de la Recherche (projects ANR-18-CE47-0004 and ANR-16-TERC-0015-01), the Conseil Regional d’Ile-de-France, Institut FRancilien des Atomes Froids, DIM Nano’K (projects METROSPIN and ACOST), DIM Sirteq (project SureSpin), and Labex FIRST-TF (project CUSAS).


\begin{thebibliography}{199}

\bibitem{Shin2008} Y. Shin et \textit{al.}, Nature \textbf{451}, 689 (2008), Y. Shin, Phys. Rev. A 77, 041603(R) (2008).

\bibitem{ho2009} T. Ho, Q. Zhou, Nature Physics \textbf{6}, 131 (2009).

\bibitem{Nascimbene2010} S. Nascimbene et \textit{al.}, Nature \textbf{463}, 1057 (2010)

\bibitem{Yefsah2011} T. Yefsah et \textit{al.}, Phys. Rev. Lett. \textbf{107}, 130401 (2011)

\bibitem{cazalilla2011} M. A. Cazalilla et \textit{al.}, Rev. Mod. Phys. \textbf{83}, 1405 (2011).

\bibitem{vortex}  M. R. Matthews et \textit{al.}, Phys. Rev. Lett. \textbf{83}, 2498 (1999), B. P. Anderson et \textit{al.},  Phys. Rev. Lett. \textbf{85}, 2857 (2000), K. W. Madison et \textit{al.}, Phys. Rev. Lett. \textbf{84}, 806 (2000), J. R. Abo-Shaeer et \textit{al.}, Science \textbf{292}, 476 (2001), E. Hodby et \textit{al.}, Phys. Rev. Lett. \textbf{88}, 010405 (2001), C. Raman et \textit{al.}, Phys. Rev. Lett. \textbf{87}, 210402 (2001), Zwierlein et \textit{al.}, Nature \textbf{435}, 1047 (2005)

\bibitem{fluctuations} R. N. Bisset, C. Ticknor, and P. B. Blakie, Phys. Rev. A \textbf{88}, 063624 (2013)

\bibitem{gross2017} C. Gross and I. Bloch, Science, \textbf{357}, 995 (2017)

\bibitem{Asteria2021}   L. Asteria et \textit{al.}, arXiv:2104.10089 (2021)

\bibitem{Subhankar2019} S. Subhankar et \textit{al.}, Phys. Rev. X \textbf{9}, 021002 (2019) 

\bibitem{Donald2019} M. McDonald et \textit{al.} Phys. Rev. X \textbf{9}, 021001 (2019).

\bibitem{wurtz2009} P. Würtz et \textit{al.}, Phys. Rev. Lett. \textbf{103}, 080404 (2009)

\bibitem{Veit2021} C. Veit et \textit{al.}, Phys. Rev. X \textbf{11}, 011036 (2021)

\bibitem{Reinaudi2007} G. Reinaudi et \textit{al.}, Opt. Lett. \textbf{32}, 3143 (2007).

\bibitem{bataille2019} P. Bataille et \textit{al.}, Phys. Rev. A \textbf{102}, 013317 (2020)

\bibitem{Menotti2002} C. Menotti, P. Pedri, and S. Stringari, Phys. Rev. Lett. \textbf{89}, 250402 (2002)

\bibitem{Sonderhouse2020} L. Sonderhouse et \textit{al.}, Nature Physics \textbf{16}, 1216 (2020)

\bibitem{bouchoule2006} J. Esteve et \textit{al.}, Phys. Rev. Lett. \textbf{96}, 130403 (2006)

\bibitem{armijo2010} J. Armijo et \textit{al.}, Phys. Rev. Lett. \textbf{105}, 230402 (2010)

\bibitem{Denschlag2000} J. Denschlag et \textit{al.}, Science, \textbf{287}, 97 (2000)



\end{thebibliography}
\end{document}